# Frequency evaluation of collimated blue light generated by wave mixing in Rb vapour

Alexander Akulshin[1], Christopher Perrella[2], Gar-Wing Truong[2], Russell McLean[1] and Andre Luiten[2,3]

[1]Centre for Atom Optics and Ultrafast Spectroscopy, Swinburne University of Technology, Melbourne, Australia
[2]School of Physics, University of Western Australia, Nedlands 6009 WA, Australia
[3]School of Physics and Chemistry, The University of Adelaide, Adelaide 5005 SA, Australia

E-mail: *aakoulchine@swin.edu.au*

**Abstract.** An evaluation of the absolute frequency and tunability of collimated blue light (CBL) generated in warm Rb vapour excited by low-power cw laser radiation at 780 nm and 776 nm, has been performed using a Fabry-Perot interferometer and a blue diode laser. For the conditions of our experiments the CBL tuning range is more than 100 MHz around the resonant frequency of the $^{85}Rb$ $5S_{1/2}(F=3) \to 6P_{3/2}(F'=4)$ transition. A simple technique for stabilizing the power and frequency of the CBL to within a few percent and 10 MHz, respectively, is suggested and demonstrated.

## 1. Introduction

The technique of frequency conversion of low-power cw radiation of diode lasers into collimated blue and far-infrared light in *Rb* and *Cs* vapours, first demonstrated by Zibrov et al [1], continues to be an active area of research [2, 3, 4, 5, 6]. Potential applications of this approach include tunable coherent light sources, quantum information processing [7] and underwater communication [8]. Key spectral characteristics of collimated blue light (CBL) such as linewidth and tuning range are important for most of these applications. Previous investigations demonstrated high temporal coherence of the CBL using a Fabry-Perot cavity and two-slit diffraction [3, 4], but did not determine the absolute frequency in detail.

CBL results from four-wave mixing (FWM) in atomic media with a diamond-type energy level configuration. An alkali vapour driven by laser radiation tuned close to strong optical transitions in a ladder-type configuration, which in the case of $^{85}Rb$ atoms are the $5S_{1/2} \to 5P_{3/2}$ and $5P_{3/2} \to 5D_{5/2}$ transitions (Fig.1a), can produce an additional optical field at 5.23 μm through amplified spontaneous emission on the $5D_{5/2} \to 6P_{3/2}$ transition. Mixing of this field and the laser fields produces optical radiation at 420 nm in the direction satisfying the phase-matching relation $k_1 + k_2 = k_{IR} + k_{BL}$, where $k_1, k_2, k_{IR}$ and $k_{BL}$ are the wave vectors of the radiation at 780, 776, 5230 and 420 nm, respectively. In contrast to conventional optical parametric oscillation, the new field generation occurs without an optical cavity. Rb atoms provide not only high Kerr nonlinearity, but also set the resonant conditions for the blue and far-IR radiation.

It is clear that the absolute frequency of the CBL must lie close to the $5S_{1/2}(F=3) \to 6P_{3/2}$ transition frequency. This is readily confirmed by the observation of isotropic blue fluorescence from Rb vapour excited only by CBL in a heated auxiliary cell. However, given that the Doppler broadening of the transition is aproximately 1 GHz, this observation does not preclude a substantial frequency detuning of the CBL from the $5S_{1/2}(F=3)\rightarrow 6P_{3/2}$ transition. From the FWM phase matching condition, the CBL frequency is

$$v_{BL}=(n_1v_1 + n_2v_2 - n_{IR}v_{IR})/n_{BL}, \qquad (1)$$

where $n_1$ and $n_2$ are the refractive indexes seen by the light fields at 780 nm and 776 nm, $n_{IR}$ and $n_{BL}$ are the refractive indexes at 5.23 μm and 420 nm, respectively, while $v_x$ is the corresponding frequency of the optical field at each wavelength. As the refractive indexes depend on the optical frequency, intensity and polarization of the applied laser radiation, as well as the atomic density $N$, the CBL frequency is a complex function of all these parameters. In this paper, we undertake an experimental study of the absolute frequency and frequency tuning range of the CBL at 420 nm generated by parametric wave mixing in Rb vapour.

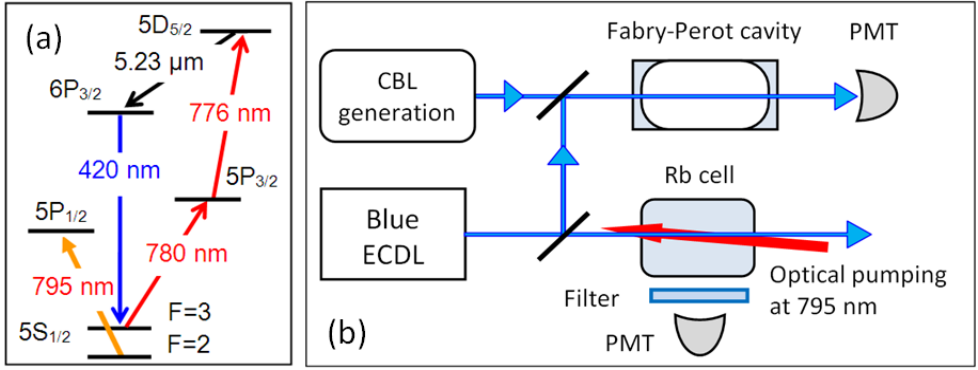

Figure 1. (a) A diagram of the [85]Rb atom energy levels involved in CBL generation and the absolute CBL frequency evaluation. (b) Schematic diagram of experimental setup.

## 2. Absolute frequency evaluation

We evaluate with high resolution the CBL absolute frequency and tuning range by measuring the frequency difference between CBL and reference radiation of known frequency. An obvious source of suitable reference radiation is a narrow-linewidth extended-cavity blue diode laser locked to a sub-Doppler feature of the $5S_{1/2}(F=3)\rightarrow 6P_{3/2}$ ($F'=2,3,4$) transition spectrum produced by standard nonlinear Doppler-free spectroscopic techniques [9]. This would readily provide a frequency accuracy of approximately 1 MHz and we could measure the frequency difference between the CBL and this radiation by comparing the spectral positions of Fabry-Perot interferometer (FPI) transmission peaks of the laser and the CBL. Unfortunately the output power of the available diode laser at 420 nm is insufficient to allow a standard saturation absorption spectroscopy arrangement and we instead implement a modified scheme for sub-Doppler spectroscopy of Rb atoms on the $5S_{1/2}\rightarrow 6P_{3/2}$ transition.

While in a conventional scheme of nonlinear Doppler-free spectroscopy both the probe and saturating beams are produced from the same laser, we use the 420 nm radiation as a probe while strong pumping radiation is provided by a 795 nm laser tuned to the *Rb D1* absorption line. This radiation produces velocity selective ground-state optical pumping of the atoms. Distinctive features of Doppler-free spectroscopy of alkali atoms under conditions in which the pump and probe radiation are produced from independent lasers and tuned to different absorption lines have previously been presented in [10].

The CBL power is enhanced or reduced according to whether the *D1* laser pumps atoms into or out of the *F=3* ground state resonant velocity group involved in the wave mixing. We could have also performed this experiment by pumping on the *Rb D2* absorption line; however, the simpler hyperfine structure of the $5P_{1/2}$ level makes the frequency comparisons easier.

Dramatic changes in the efficiency of CBL generation by simultaneously driving the *D1* transition have been demonstrated in [6], and here we use a similar idea for producing sub-Doppler frequency references for the blue laser.

We note also that the use of commercially available optical spectrum analysers and wavemeters to determine the absolute frequency of the CBL are not appropriate as they require an absolute frequency calibration and do not provide the spectral resolution needed for the frequency determination: the typical resolution of wavemeters based on a Fizeau etalon is approximately tens of GHz.

### 3. Experimental set-up

Our experimental setup, schematically illustrated in Fig. 1b, is similar to that used in previous experiments [3, 6]. Radiation from extended cavity diode lasers (ECDLs) is used for step-wise and two-photon excitation of Rb atoms at 780 and 776 nm. The frequency of the 780 nm ECDL can be swept across the *D2* absorption line or modulation-free stabilized using the Doppler-free polarization resonance obtained on the $^{85}$*Rb* $5S_{1/2}(F=3) \to 5P_{3/2}(F'=4)$ transition in an auxiliary *Rb* cell. The 776 nm laser is also swept across the $^{85}$*Rb* $5P_{3/2} \to 5D_{5/2}$ transition or modulation-free locked to a low-finesse tunable confocal etalon. Typical values of the standard deviation of the frequency noise for both lasers in the frequency stabilized regime are in the range 200 - 300 kHz over a 1 second time interval, while the laser linewith is approximately 1 MHz.

Radiation from both ECDLs is combined to form a bichromatic beam. The powers of the components at 780 and 776 nm are typically 12 and 6 mW, respectively. The bichromatic beam is focused into a 5 cm long heated Rb cell containing a natural mixture of Rb isotopes with no buffer gas. The cross section of the beam inside the cell is about 0.5 mm$^2$. The temperature of the cell is set within the range 60-100$^\circ$C meaning that the density *N* of *Rb* atoms varies from $3 \times 10^{11}$ cm$^{-3}$ to $6 \times 10^{12}$ cm$^{-3}$.

An additional Rb cell heated to 60$^\circ$C and two ECDLs tuned to the $^{85}$*Rb D1* line and to the $5S_{1/2} \to 6P_{3/2}$ transition at 420 nm are employed to determine the absolute frequency reference for the CBL.

Colour filters with optical density approximately 0.5 and 4.0 at 420 nm and 780 nm, respectively, are used to spectrally select the CBL and isotropic blue fluorescence, both of which are detected by photomultipliers (PMTs). A μ-metal shield is used to reduce the ambient magnetic field in the *Rb* cell to a few milligauss.

The CBL spectral purity and linewidth are explored using a tuneable concave mirror FPI of length *L*=14.5 cm and having high finesse in the blue spectral region. The spatial distribution of the blue light transmitted through the interferometer is monitored with a CCD camera to ensure that the radiation is mainly coupled into the fundamental TEM$_{00}$ interferometer mode by a combination of lenses.

### 4. Results

*4.1. Spectral properties of the CBL*

Figure 2 shows the CBL intensity profile and blue FPI transmission resonances, obtained at intermediate atomic density $N \approx 1.2 \times 10^{12}$ cm$^{-3}$, as a function of the 776 nm laser frequency, while the 780 nm laser is locked to the $5S_{1/2}(F=3) \to 5P_{3/2}(F'=4)$ cycling transition. The CBL doublet structure is laser intensity and frequency dependent, as has been discussed in [3]. The frequency range of the 776 nm laser over which the CBL generation occurs is considerably smaller than the Doppler width of the $^{85}$*Rb D2* line at this temperature. Curves (*ii*) and (*iii*) represent the blue light transmission through the fixed-length FPI and demonstrate that CBL can be coupled mainly into the fundamental TEM$_{00}$ modes of the interferometer. The blue FPI has been tuned so that transmission peaks of the axial modes approximately coincide with one of the maxima of the CBL profile. Narrow, high-contrast transmission resonances for various FPI tunings over the entire CBL profile confirm the single-frequency narrow-linewidth spectrum of the CBL. It is important to appreciate that the approximately 400 MHz tuning range of the 776 nm laser over which CBL is observed in Fig. 2 does not imply that the CBL frequency itself is varying by that amount. Tuning the 776 nm laser is also expected to cause the frequency of the 5.23 μm light to vary for example, and we cannot directly detect this variation.

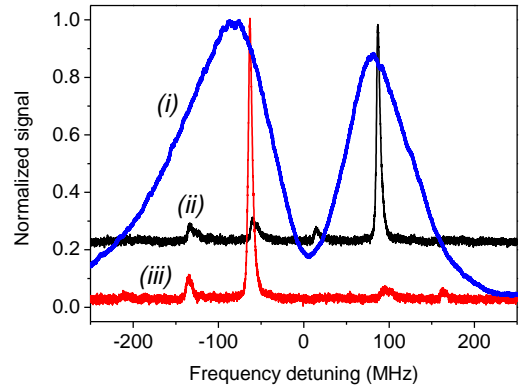

Figure 2: The CBL profile (*i*) and blue FPI transmission resonances (*ii*) and (*iii*) vs. frequency of the 776 nm laser scanned across the $^{85}Rb$ $5P_{3/2} \rightarrow 5D_{5/2}$ transitions, while the 780 nm laser is locked to the Doppler-free resonance on the $^{85}Rb$ $5S_{1/2}(F=3) \rightarrow 5P_{3/2}(F'=4)$ cycling transition. Zero detuning of the 776 nm laser is chosen arbitrarily to coincide with the CBL minimum and does not necessarily imply zero detuning from the 776 nm transition. The baseline of curve (ii) is moved vertically for clarity.

## 4.2. *Blue laser frequency evaluation*

We now consider how the velocity-selective optical pumping technique can be used in evaluating the absolute frequency of the blue ECDL.

If the fixed frequency 420 nm laser is tuned to the inhomogeneously broadened $^{85}Rb$ $5S_{1/2}(F=3) \rightarrow 6P_{3/2}$ transition, then the main contribution to the fluorescence signal comes from atoms excited on the strongest cycling transition $5S_{1/2}(F=3) \rightarrow 6P_{3/2}(F'=4)$ at frequency $v_{34}$ and having velocity component $v_z = 2\pi(v_{BL} - v_{34})/k_{BL}$ in the direction of the laser beam. Two other resonant velocity groups which interact on the weaker open transitions $5S_{1/2}(F=3) \rightarrow 6P_{3/2}(F'=2, 3)$ are significantly pumped into the *F=2* ground-state level. The population of the $v_z = 2\pi(v_{BL} - v_{34})/k_{BL}$ velocity group in the *F=3* ground-state level could also be modified by hyperfine optical pumping if it also interacts with additional radiation. This results in resonances in the blue fluorescence signal if, for example, the monochromatic counter-propagating pumping radiation at 795 nm is detuned by $\delta v_{D1} = (v_{ij} - v_{D1}) = k_{D1}v_z/2\pi$ from either transition within the $^{85}Rb$ D1 line, where $v_{ij}$ is the resonant frequency of the $5S_{1/2}(F=i) \rightarrow 5P_{1/2}(F'=j)$ transition. Thus, the blue laser detuned from the $5S_{1/2}(F=3) \rightarrow 6P_{3/2}(F'=4)$ transition by an amount $\delta v_{BL} = (v_{BL} - v_{34})$ produces an enhanced fluorescence signal if the pumping laser frequency is detuned from any transition starting from the *F=2* ground-state level. We note, however, that the detuning $\delta v_{D1}$ differs from $\delta v_{BL}$ because of the large frequency difference between the blue and near-IR radiation:

$$\delta v_{D1} = v_{2j} - (v_{BL} - v_{34})(k_{BL}/k_{D1}) \approx v_{2j} - 1.86 \times \delta v_{BL} . \qquad (2)$$

Thus, we are able to tune the frequency of the blue laser precisely to the $^{85}Rb$ $5S_{1/2}(F=3) \rightarrow 6P_{3/2}(F'=4)$ transition through a comparison of the spectral positions of blue fluorescence resonances and saturated absorption resonances on the $5S_{1/2}(F=2) \rightarrow 5P_{1/2}(F'=2; 3)$ transitions.

Figure 3 shows normalized fluorescence at 420 nm produced by the fixed-frequency $v_{BL}$ blue laser and plotted as a function of the pumping laser frequency $v_{D1}$ as it is swept across the inhomogeneously broadened $^{85}$Rb $5S_{1/2}(F=2) \rightarrow 5P_{1/2}(F'=2; 3)$ transitions. In Fig. 3a the fluorescence is enhanced above the no-pumping level at four different frequencies of the pumping laser rather than the expected two. The two smaller peaks arise from optical pumping produced by 795 nm light back-reflected from the cell window. The absolute frequency of the optical pumping laser can be estimated from standard saturated absorption resonances observed in an auxiliary cell. The frequency offset of the blue fluorescence resonances from these sub-Doppler saturated absorption resonances on the *D1* line allows the blue laser detuning from the $5S_{1/2} \rightarrow 6P_{3/2}$ transitions to be estimated using equation (2). For example, the frequency shift shown in Fig.4a suggests that the blue laser is tuned approximately 60 MHz above the $5S_{1/2}(F=3) \rightarrow 6P_{3/2}(F'=4)$ transition frequency. If the blue laser is tuned precisely on resonance, the large and small fluorescence peaks merge and spectrally coincide with the saturated absorption resonances on the *D1* line, as shown in Fig. 3b. In this condition, the 420 nm laser is resonant with the atoms that have zero axial velocity. Figure 3b also illustrates power broadening of the sub-Doppler resonances. It is possible to obtain better resolution with lower pumping power.

The blue fluorescence produced by the 420 nm laser can be reduced rather than enhanced if the pumping laser depletes the resonant velocity group by pumping to the uncoupled ground hyperfine level. This is demonstrated in Figure 4 where the fluorescence resonances with sub-Doppler width coincide spectrally with saturated absorption resonances on the $5S_{1/2}(F=3) \rightarrow 5P_{1/2}(F'=2; 3)$ transition. We estimate that the blue diode laser is tuned to the $^{85}$Rb $5S_{1/2}(F=3) \rightarrow 6P_{3/2}(F'=4)$ transition with approximately ±5 MHz accuracy. The fluorescent signals shown in Fig. 5 have been power broadened because of the high-power pumping radiation used in this case.

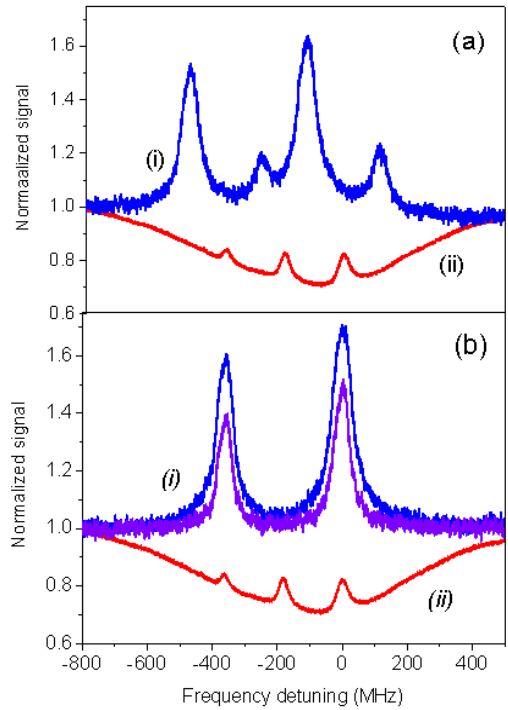

Figure 3: Blue fluorescence (curves *i*) as a function of pumping laser detuning from the $^{85}$Rb $5S_{1/2}(F=2) \rightarrow 5P_{1/2}(F'=3)$ transition for different frequencies of the blue laser normalized on the fluoresence signal without optical pumping: (a) large frequency detuning (~60 MHz) of the blue laser from the $5S_{1/2}(F=3) \rightarrow 6P_{3/2}(F'=4)$ transition at a pumping radiation intensity of 25 mW/cm$^2$; (b) blue laser tuned to the $5S_{1/2}(F=3) \rightarrow 6P_{3/2}(F'=4)$ transition at pumping intensities of 25 and 8 mW/cm$^2$. Curves (*ii*) are saturated absoption profiles on the $^{85}$Rb $5S_{1/2}(F=2) \rightarrow 5P_{1/2}$ transition obtained in the axillary *Rb* cell kept at ~50 °C.

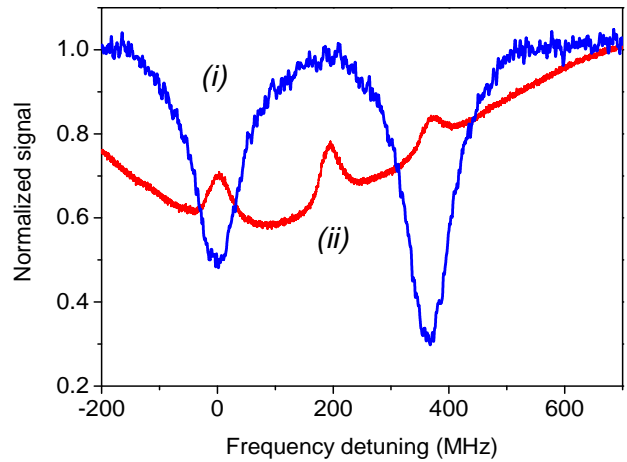

Figure 4: Normalized blue fluorescence excited by the fixed frequency blue laser tuned to the $^{85}Rb$ $5S_{1/2}(F=3) \rightarrow 6P_{3/2}(F'=4)$ transition and saturated absorption resonances as a function of the pumping laser detuning from the $^{85}Rb$ $5S_{1/2}(F=3) \rightarrow 5P_{1/2}(F'=2)$ transition.

Figure 5 shows Fabry Perot cavity transmission resonances of the CBL generated in the *Rb* cell and the 420 nm laser tuned to the $5S_{1/2}(F=3) \rightarrow 6P_{3/2}(F'=4)$ transition using the procedure described above. It is seen that the peaks are separated by approximately 25 MHz in this figure, but this can easily be changed by tuning the frequencies of the 776 nm and 780 nm lasers around their corresponding transitions. In this way we find that the CBL intensity is maximized when it is tuned precisely on the $5S_{1/2}(F = 3) \rightarrow 6P_{3/2}(F'=4)$ transition, and that the CBL peak can be tuned by more than 100 MHz around this transition frequency. This is the most direct way we have of determining the tuning range of the CBL.

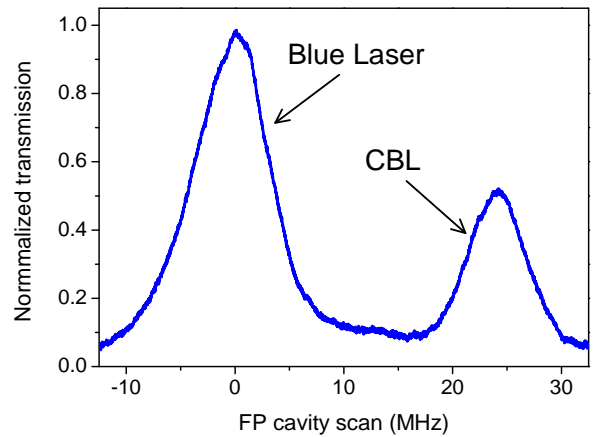

Figure 5: Fabry Perot interferometer transmission resonances while all laser frequencies are fixed. The 420-nm laser is tuned to the $^{85}Rb$ $5S_{1/2}(F=3) \rightarrow 6P_{3/2}(F'=4)$ transition with accuracy of ±5 MHz. (Resonances are broadened due to signal averaging.)

*4.3. CBL frequency stabilization*
For the CBL to be used effectively as a narrow-band light source, it is desirable to have a means of stabilizing its frequency and amplitude. As has been discussed above, the frequency of the CBL depends primarily on the frequencies of the two applied laser fields. Assuming the other factors are not varying, stabilizing the frequencies of the two applied laser fields can stabilize the CBL frequency and to some extent the CBL power. While the 780 nm laser can be locked routinely to the strong $5S_{1/2}(F=3) \rightarrow 5P_{3/2}(F'=4)$ cycling transition with standard Doppler-free spectroscopic techniques, this is not the case for the 776 nm laser as it is tuned to a transition between excited states. Although one could stabilize this laser to a reference cavity, a more elegant solution is to lock the 776 nm laser directly to the CBL profile.

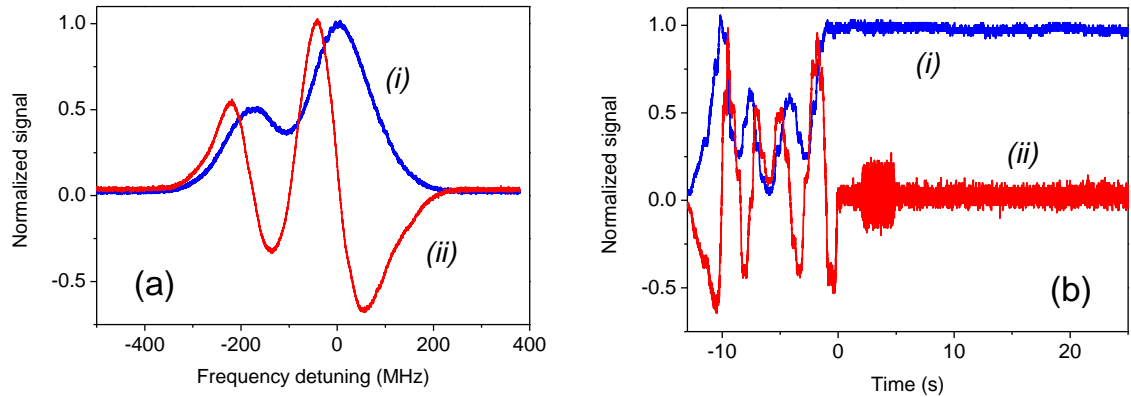

Figure 6: CBL peak locking. (a): Curves *(i)* and *(ii)* show CBL intensity profile and its first derivative, respectively, resulting from the 776 nm laser scan while the 780 nm laser is locked to the $5S_{1/2}(F=3) \rightarrow 5P_{3/2}(F'=4)$ transition.
(b): Curves *(i)* and *(ii)* show CBL intensity and error signal temporal variations recorded with slow manual frequency scanning of the 776 laser (*t*<0) and in frequency locking regime (*t*>0).

Here we demonstrate that both peak and side locking can be used to control the step-wise excitation of Rb atoms and, consequently, the CBL frequency and intensity, with the error signal derived from the CBL intensity itself.

For the peak locking, the dither voltage at 10 kHz from a lock-in amplifier modulates the 776 nm laser frequency. The lock-in amplifier output, which is proportional to the first derivative of the CBL spectral profile, is used as an error signal. Curve (*i*) in Figure 6a shows the intensity profile of the CBL as the 776 nm laser is scanned across the $5P_{3/2} \rightarrow 5D_{5/2}$ transition, with the 780 nm laser locked to the $^{85}$Rb $5S_{1/2}(F = 3) \rightarrow 5P_{3/2}(F' = 4)$ transition. Curve (*ii*) is the derivative signal. As a modest modulation index is applied, the broadening of the CBL reference line due to the laser modulation is small compared to the FWHM of the entire profile.

Active stabilization of the CBL is shown in Figure 6b. The 776 nm laser is initially scanned manually, then tuned to the higher peak of the CBL profile and the locking loop closed at *t=0*. The error signal enhancement that occurs in the *2s<t< 5s* interval is due to servo system gain variations in a search for optimum values of the locking parameters.

Assuming that in the locked mode the error signal fluctuations correspond to CBL frequency variations, we estimate that the relative frequency stability after locking is $\Delta \nu_{BL}/\nu_{BL} < 8.0 \times 10^{-10}$. From curve *(i)* we find that in the locked case the relative CBL power fluctuations are less than 2%.

It is worth noting that CBL self locking is very robust due to the large recapture range provided by the relatively wide reference line. The considerably improved long-term frequency and intensity stability demonstrates the effectiveness of this simple method.

We have also demonstrated that the 776 nm laser can be locked to the side of the CBL resonance, eliminating the need to dither the 776 nm laser frequency and preserving the intrinsic linewidth of the

CBL. Even higher values are obtained for the relative frequency stability of the CBL ($\Delta\nu_{BL}/\nu_{BL} \approx 6\times10^{-10}$ for 10 s); however, the coupling between CBL power and frequency variations and sensitivity to beam alignment impose limits on the long-term stability achievable with this approach. Side locking is preferable if high coherence of the CBL is required.

We note also that a similar arrangement for locking the sum frequency of two lasers to Doppler-free fluorescence resonances that result from two-photon excitation of Rb atoms on the $5S_{1/2} \rightarrow 5D_{5/2}$ transition was implemented in [11]. In that work the reference resonances are much narrower, but more importantly their absolute frequency is determined by the separation between the Rb energy levels, while the CBL frequency is a complex function of the parametric FWM process.

## 5. Conclusion

We have investigated the absolute frequency of collimated blue light generated by a parametric four-wave mixing process in an atomic Rb vapour. The frequency of the CBL has been compared with the frequency of a blue diode laser using a tunable Fabry-Perot interferometer. The blue laser frequency has, in turn, been determined using a modified sub-Doppler spectroscopy technique where velocity-selective pumping radiation has been applied on the *Rb D1* line. This has allowed the absolute frequency of the CBL to be evaluated with approximately ±5 MHz precision, limited primarily by the precision with which we are able to determine the frequency of the laser. The CBL frequency is found to be centred on the $5S_{1/2}(F=3) \rightarrow 6P_{3/2}(F'=4)$ transition, and can be tuned over a range of more than 100 MHz around this frequency by tuning the frequencies of the driving lasers.

Finally, we have proposed and demonstrated a method for stabilizing the power and frequency of the CBL, in which an error signal is derived from the CBL profile itself and used to lock the frequency of the 776 nm laser involved in the wave mixing process.

**Acknowledgments**

This work has been supported by the ARC Centre of Excellence for Quantum-Atom Optics. AL, CP and GWT would like to thank the ARC for supporting this research through the DP0877938 and FT0990301 research grants.